**Title**
Designing optimal scaffold topographies to promote nucleus-guided mechanosensitive migration using *in silico* models

Maxime Vassaux[1]*, Laurent Pieuchot[2,3], Karine Anselme[2,3], Maxence Bigerelle[4] and Jean-Louis Milan[1]*
1 Aix Marseille Univ, CNRS, ISM, Marseille, France;
2 Université de Haute-Alsace, CNRS, IS2M, UMR 7361, Mulhouse, France;
3 Université de Strasbourg, Strasbourg, France; and
4 Université de Valenciennes et du Hainaut Cambrésis, Laboratoire d'Automatique, de Mécanique et d'Informatique industrielle et Humaine (LAMIH), UMR-CNRS 8201, Le Mont Houy, Valenciennes, France

*Correspondence: jean-louis.milan@univ-amu.fr or m.vassaux@ucl.ac.uk

**Abstract**

Computational models have become an essential part of exploratory protocols in cell biology, as a complement to *in vivo* or *in vitro* experiments. These virtual models have the twofold advantage of enabling access to new types of data and validate complex theories. The design of mechanically functionalized biomaterials or scaffolds, to promote cell proliferation and invasion in the absence or in the complement of synthetic chemical coatings, can certainly benefit from these hybrid testing approaches. The underlying fundamental process of cell migration and in particular its dependence on the cell mechanical/geometrical environment remains crudely understood. Currently at least two theories explain the migration patterns observed by cells on curved topographies, involving either polymerization dynamics of actin or assembly dynamics of focal adhesions. We recently proposed a third mechanism relying on nucleus mechanosensitivity, which has been tested extensively experimentally and computationally. We now explore the hypothesis that nucleosensitivity could be a mechanism for cells to optimally find microenvironments suited for mitosis, providing mechanical stability and relaxation. By means of a computational mechanical model with intracellular structure detail, we investigate how the complex interplay between this new migration mechanism and the microenvironment topography can lead to more relaxed cells and organelles. To go further, we simulated in this study cell migration via a novel protocol in silico which generates dynamical ripple wave on a deformable substrate and changes topography over time. This kind of in silico protocols based on a new understanding of cell migration and nucleosensitivity could, therefore, inform the design of optimized scaffold topographies for cell invasion and proliferation.

# 1. Introduction

## 1.1. Intertwined computational-experimental protocols

Computational physical models are large sets of equations that describe a controlled, reduced version of an experiment. Unlike analytical models, computational ones can integrate a more significant part of the complexity of living systems, as computational resources and methods allow to solve numerous and complex equations on large and heterogeneous systems. Nevertheless, *in silico* models remain far from the full complexity of *in vivo* experiments, and their validity relying on various assumptions can always be questioned. *In vitro* models are a first step toward breaking down the physics of living systems, disentangling that complexity, but *in silico* ones constitute a step further. Hypothesized multiple physics and the multiple scales involved in the mechanisms regulating the behavior of living systems can easily be integrated and tested as desired within *in silico* models. They have become an essential tool in theoretical biophysics, complementary to *in vivo* and *in vitro* models [Mogilner09, Rodriguez13, Rens17]. Indeed, *in silico* models enable integrating more complexity in a controlled way. For example, in discerning the origin of observed biological behavior, *in silico* models enable to sort active regulatory mechanism from passive physics [Nickaeen19, Winkler19]. *In silico* models also provide complementary data, hardly accessible with *in vitro* and even less *in vivo* models: (i) at different scales, very small ones, for example, using methods solving the mechanics of clouds of electrons [Zink13] (ii) and of different type, quantities that cannot be measured directly, called internal state variables in thermodynamics, among which forces and stresses can be found for example [Milan16].

### 1.2. Biomaterials design for tissue engineering

The design of biocompatible materials for tissue engineering requires understanding how cells and materials interact to promote cell proliferation and invasion. Cell biology and more specifically migration are conditioned by the scaffold (or substrate) physics, which constitute a set of cues relying on chemistry [Zigmond73, Dillon95], electromagnetics [Adley83] and mechanics [Isenberg09]. These physics play a role on multiple scales predominantly ranging from the characteristic scale of electrons, up to at least the characteristic scale of the cell. However, it is not possible yet to exclude larger scales, as macroscopic thermodynamical effects could certainly influence cell mechanobiology [Isenberg09]. Unravelling the interplay between these different cues at different scales and cell migration would certainly enable to engineer optimal biomaterials.

### 1.3. The example of MAPS: could topography cause invasion?

Recently developed biocompatible scaffolds made of a microporous annealed particle (MAP) gel display promising levels of cell invasion and proliferation [Griffin15, Darling18]. MAP gels are an example of biomaterial making use of the newly understood cell signaling cues. The mechanical properties of the gel can be tuned to steer the differentiation of cells depending on the type of tissue to repair Annealing particles of controlled sizes also enable to tune stability and therefore biodegradability of the scaffold. Nevertheless, the higher orders of cell invasion and proliferation are hardly explained by these features of the gels. More generally, microporous scaffolds tend to induce similar enhanced cell migration rates. In the meantime, recent pieces of research have highlighted *in vitro* and confirmed *in silico* how the geometry and the topography of the substrate can direct migration [Clark91, Doyle09, Czeisler16]. Specifically, the mechanical instability conveyed by convex topographies [Vassaux17, Pieuchot18] is shown to promote cell motility [Vassaux19]. Could a link more substantial than a correlation between the observed enhanced invasion and the topographical cue be established here? Potentially, causation involving underlying cell mechanics?

## 2. Understanding cell migration in interaction with extracellular topography

### 2.1. Current theories and in silico models used to explore them

Cells have evolved multiple mechanisms to migrate in interaction with their environment. For instance, we reported recently a new cellular ability, which we termed "curvotaxis" that enables the cells to respond to cell-scale curvature variations, a ubiquitous trait of cellular biotopes [Pieuchot18]. Ascertaining enhanced invasion as a consequence of the specific geometry of microporous scaffolds lacks a mechanistic explanation of how cell migration is systematically influenced by the curvature of the underneath substrate. Topography and curvature influence cell physics provoking local confinement at cell-scale or below, from which originates cell polarization. At least three theories have been developed in the last two years and supported by means of computational models [Winkler19, Vassaux19, Schakenraad19]:

### a) Topography as confinement of actin polymerization during cell migration

Winkler et al. (2019) have shown that confinement breaks down the symmetry of actin polymerization in the cytoskeleton, and therefore favors a particular direction of the extension of the lamellipodium. This mechanism is endorsed by simulating physiological migration patterns using a continuum phase-field model of a single adherent cell and its internal actin organization. Actin ordering and polymerization seen as key factors of cell motility is not a recent discovery [Mogillner09], however, as a source of persistence in confined environments definitely is. Even pieces of evidence in epithelia show a correlation between actin organization and topography, actin flowing away from parts of the cytoskeleton exposed to convex curvatures [Chen19].

### b) The nucleus pushed away from convex topography indicates the direction for cell migration to more relax region

We have rather focused on the role of the nucleus in [Vassaux19]. Confinement is shown to polarize the nucleus position inside the cell and we hypothesized that nucleus internal motility is a precursor of migration guidance. We had recently proved a correlation between the direction of nucleus motility and the direction of cell migration on sinusoidal surfaces (Fig. 1) [Pieuchot18]. We simulated cell adhesion on sinusoidal surfaces using our particle-based model of a single adherent cell with an explicit description of the nucleus [Vassaux17, Adhsc18]. Simulations indicated a decentering of the nucleus toward the valleys of the sinus, which are concave regions of lower pressure (Fig. 2). We integrated secondly the mechanism of cell migration in the direction of intracellular nucleus displacement. Cell model reached concave regions whatever is its initial deposit location (Fig. 3). Simulations of persistent migration away from convex topographies and stabilization on symmetric concave niches supported our theory. This nucleus mechanosensitive mechanism could explain the intensive cell invasion and proliferation observed in MAP scaffolds, in which cells are solely exposed to convex surfaces.

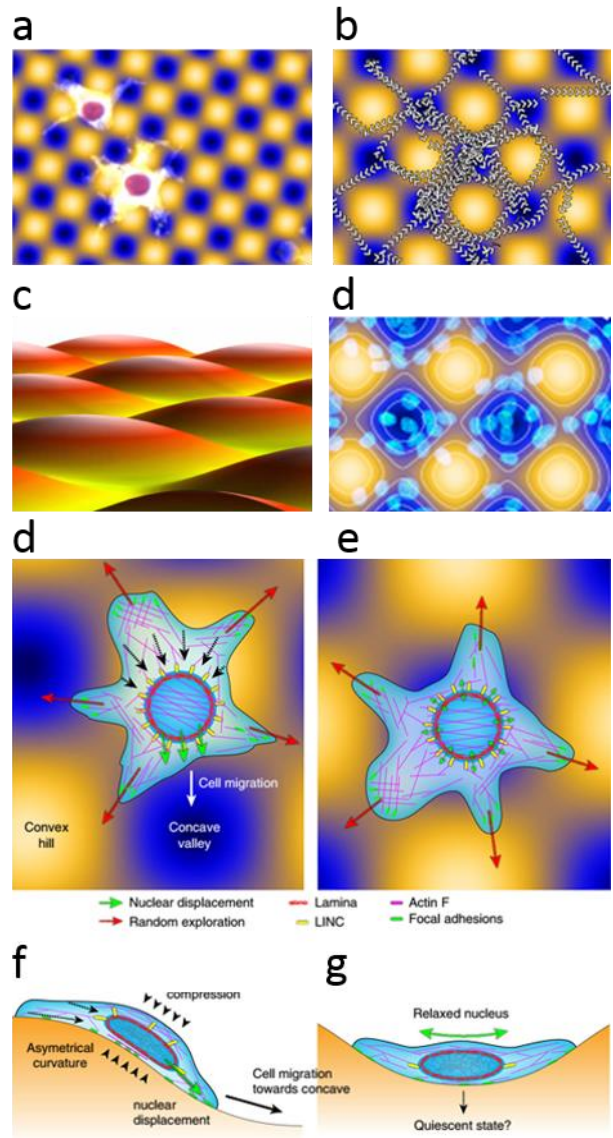

*Figure 1: in vitro observations of cell cultured on surfaces with sinusoidal topography (a). Cell migration trajectories remain in valleys and avoid peaks (b) of the sinusoidal topography viewed by side (c). Later, cells tend to stabilize their position in concave regions (d). Mechanical hypothesis to explain cell migration on sinus: curvature gradient breaks homogeneity in the compressive stress exerted by the* cytoskeleton *on the nucleus (f). The nucleus moves to lower pressure region. Then the cell migrates in the direction of nucleus movement so that the nucleus is in the center of the cell (g). Reproduced under the terms of the CC-BY 4.0 license [Pieuchot18].*

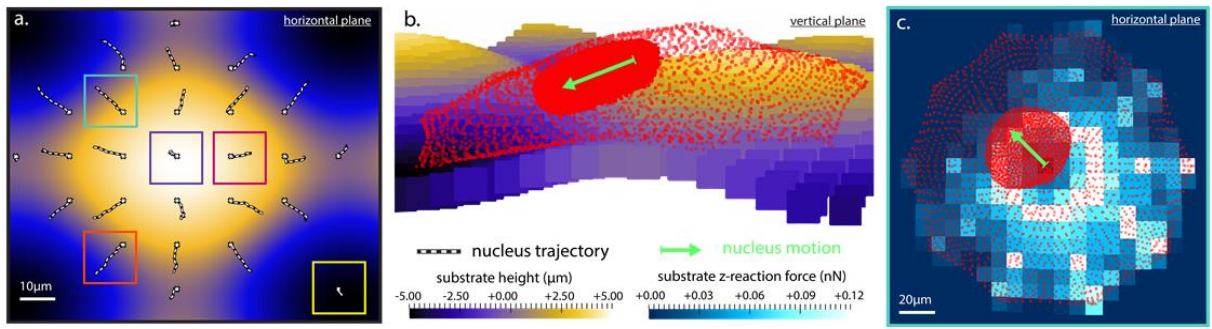

*Figure 2: Final displacements of the nucleus in the cell model depending on cell location on the sinus (a). Nucleus motion in cell model adhering on a peak (b) Nucleus motion in the opposite direction of greater traction force on the substrate. Reproduced under the terms of the CC-BY 4.0 license [Vassaux19].*

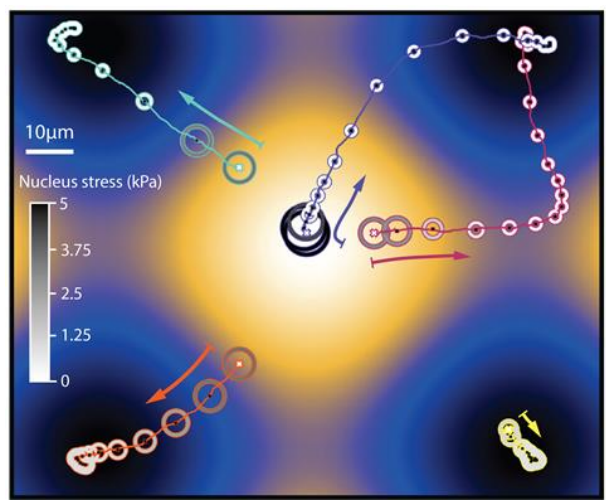

*Figure 3: Simulation of cell migration following the curvature-induced nucleus displacement. The cell model stabilizes when the nucleus stabilizes, the both in the center of the concave region. Reproduced under the terms of the CC-BY 4.0 license [Vassaux19].*

   c) Cell migration between obstacles as brownian particle movement involving repelling force

Schakenraad et al. (2019) have actually led an investigation of cell migration one scale above, to which the cell is modelled as an active Brownian particle, assuming that the influence of the environment's topography is cell-type independent. Cells consist of rigid disks with a finite area, imposed with a self-propelling velocity. The magnitude of the imposed velocity is constant, and its direction is defined as fluctuating randomly with some persistence. Cell migration is simulated on a substrate paved with obstacles which simply exert a repelling force modifying the overall cell motion, but not the imposed velocity. Schakenraad et al. (2019) have shown that the observed guidance of migration on such substrates could be caused entirely by the spatial modulation of obstacles, cells crudely migrating toward less confined spaces, independently organelles mechanosensitivity.

These 3 theories we presented hereabove are all valid, have redundancies and grey areas, therefore not mutually exclusive, but need to be sorted. These theories illustrate separate mechanisms of guided migration in anisotropic environments, at the scale of the organelle or of the whole cell. The actin cortex contractility is probably a redundant element of the two first theories, as it builds up the pressure gradient in the cytoskeleton polarizing either actin polymerization or nucleus position. In turn, cell migration may be, respectively, either a passive mechanism or a nucleus centering regulation mechanism [Almonacid15].

### 2.2. How conceptually mechanical in silico cell models can be used at the interface between materials science and cell biophysics for scaffold design?

Harnessing cells polarization and migration mechanisms, the design of scaffolds could be improved. The scaffold topography and the induced confinement could be tuned to promote optimal motility depending on the cell lineage and its characteristic mechanical properties. *In silico* mechanical cell models constitute a tool of choice for scaffold design. Such models integrate simultaneously the topographical and mechanical complexity of the cell microenvironment that is the scaffold and the mechanosensititvity of the cell migration process. In turn, sensitivity analysis of migration rate and persistence to the scaffold design parameters is rendered easily tractable.

## 3. Going deeper in the understanding of cell migration with an *in silico* cell model

### 3.1. Description of the mechanical in silico cell model

The *in silico* cell model we are developing integrates substrate and cell dynamics describing the mechanical structure as assembly of rigid particles [Vassaux17]. The model explicitly integrates actin, microtubules, intermediate filaments networks, contractile stress fibers, a contractile actomyosin cortex mingled in the cytoplasmic membrane, a viscous cytosol, and a viscoplastic nucleus (Fig 4.a). Each internal cell structure is modelled as an assembly of particles interacting via contact or springs. The parameterization of the model's interaction potentials has been largely verified and validated against indentation tests on mesenchymal stem cells [Vassaux17]. Complete details on the mathematical foundation of the model as well as the calibration, validation, and adhesion simulation process can be found in [Vassaux17]. This mechanical cell model is able to capture realistic nucleus dynamics; the nucleus equilibrium is found at the center of the cell on a flat topography. These are governed by the coupled contribution of viscous, inertial (nucleus mass), and elastoplastic (conformational changes in the cytoskeleton) effects.

Simulations of cell adhesion follow a standardized procedure. In their initial configuration, the simulated cells display a spherical shape (Fig 4.b). Spreading is actioned after the displacement of the focal adhesions (FAs) away from the center of the cell following the topography of the substrate (Fig 4.c). This dynamic adhesion process, coupled with actomyosin contraction in stress fibers and the actin network, induces conformational changes in the cytoskeleton. At the end of the simulation (Fig. 5), cells are pulled onto the substrate and attached via a set number of focal points. FAs are distributed at the cell's periphery, regardless of the site of the cell adhesion in a concave, convex, or in the transitional areas.

Subsequently, the adhesion model has been extended to render migration tractable [Vassaux19]. The migration is simulated by reproducing in a simplified way the simultaneous protrusion of a lamellipodium at the front and the cell retraction at the back of the cell. The cell model migrates as new FAs are continuously assembled away from existing disassembling adhesions in the direction of motion. While the cytoskeleton connects the new FAs, the old ones are disassembled. We hypothesized that the lamellipodium forms in the direction of the topography-induced polarization of the nucleus and advances proportionally to nucleus internal motion. The internal displacement of the nucleus is computed as the vector directed from the cell barycenter to the nucleus barycenter. The spatial jump (amplitude, direction) from the disassembled adhesions to the assembled ones at a given step is equal to the internal displacement of the nucleus observed at the previous step. The simulation of cell migration ends when the nucleus displacement becomes negligible with respect to the cell dimensions; that is when the cell is assumed to have stabilized. Such a procedure renders a continuous migration of the cell.

The level of complexity encompassed in such *in silico* model is already high enough so that we are enabled to investigate the role of several intracellular structures, as well as the topography of substrate on cell mechanics, as well as on a hypothesized nucleosensitive migration mechanism. In comparison to *in vitro* models, the mechanical properties of each component of the model may easily be tuned and their role assessed on migration parameters such as rate and persistence. Acquiring data is also simplified, as in such computational models, mechanics are intrinsically quantified.

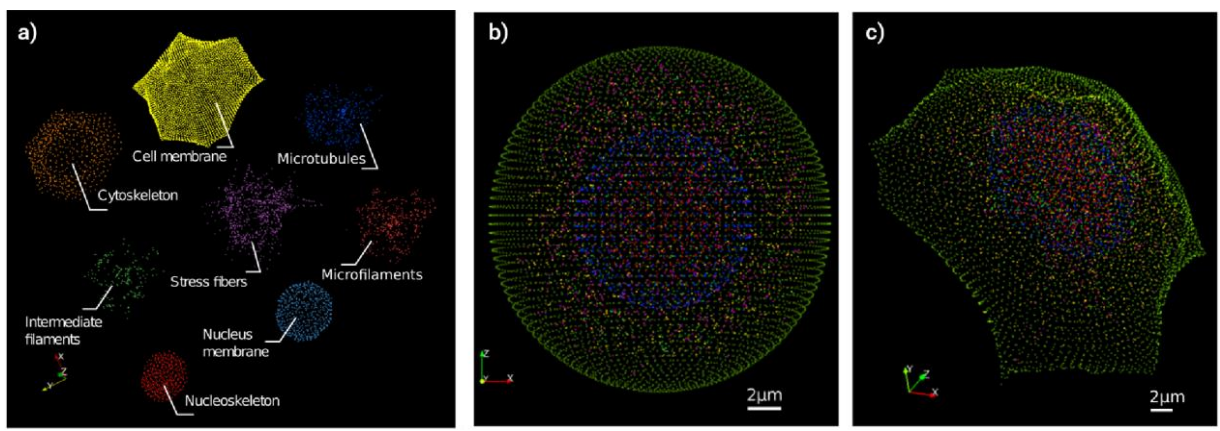

*Figure 4: Detailed structure of our in silico mechanical cell model. In the cell model, particles are interacting in one of three ways: repulsive contact, cable-like or spring-like. (a) The model encompasses a wide range of intracellular structures, for more details on what type of interaction is used for each structure and why see [Vassaux17]. (b) Initially, before adhesion, the cell is generated in a spherical shape, with all its constituents relaxed. (c) After adhesion, here on a convex substrate (not appearing), the cell finds its stretched configuration, with the filaments and stress fibres building up tension, and the microtubules and nucleus bearing compressive loads, ensuring structural stability of the cell. Reproduced under the terms of the CC-BY 4.0 licence [Vassaux17].*

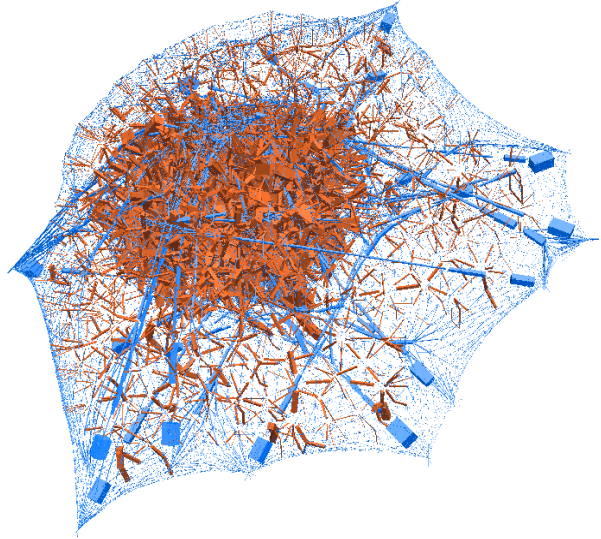

*Figure 5: Intracellular force network in the in silico cell model. Blue and red segments represent respectively tension and compression forces. Width of the segment is proportional to the magnitude of the force.*

### 3.2. Cell-scale curvatures optimize migration rates and persistence

We hypothesized the importance of wavelength and amplitude for the nucleosensitivity guidance mechanism to occur. We made use of our *in silico* cell model whereby cell motility is induced by direction and the magnitude of the polarization of the nucleus to find optimal sinusoids to promote single mesenchymal stem cell migration rate. We demonstrated that on cell-scale curvatures an optimum of migration efficiency is reached. Cells were arbitrarily positioned in the neutral part of the sinusoid, that is in the middle of a flat portion of the sinusoid where the curvature is null. The adhesion and migration dynamics were simulated on three sinusoids, with a constant ratio of amplitude to wavelength: 3µm/30µm, 10µm/100µm, 30µm/300µm (Fig. 5). The dynamics were observed until the cells stabilize and their motile behavior was considered inexistent. The efficiency of the cell model in finding the direction of the shortest path to the location of stabilization varied significantly with the sinusoid size. Radii of curvature approximately of the size of the cell led to the most straightforward to stabilization. On shorter and larger radii of curvature, cells exhibited curved trajectories (Fig. 5.a) or even sudden changes of direction (Fig. 5.c). In turn, migration rates were also much higher on cell-scale curvatures, reducing the time from the initiation of the dynamic migratory behavior to stabilization.

Simulations indicated that curvotaxis at small wavelength seems limited. Similarly, long-wave curvotaxis is also limited: large sinusoids are almost flat surfaces that offers almost no relaxation zone. The cell model cannot sense larger wavelength than its own spread diameter nor it senses smaller wavelength than the diameter of its nucleus. In other words, in this mechanism of cell migration induced by curvature and intracellular movement of the nucleus, the diameter of the nucleus and the diameter of the cell constitute the minimum and the maximum of the spatial scale of the curvotaxis of the cell ; cell curvotaxis is then related to its intrinsic dimensions. Extrapolating these results to scaffold design, topographies exhibiting cell-scale curvatures could be used for enhanced invasion and proliferation. As cells stabilize faster, they also enter more rapidly in growth and division phase.

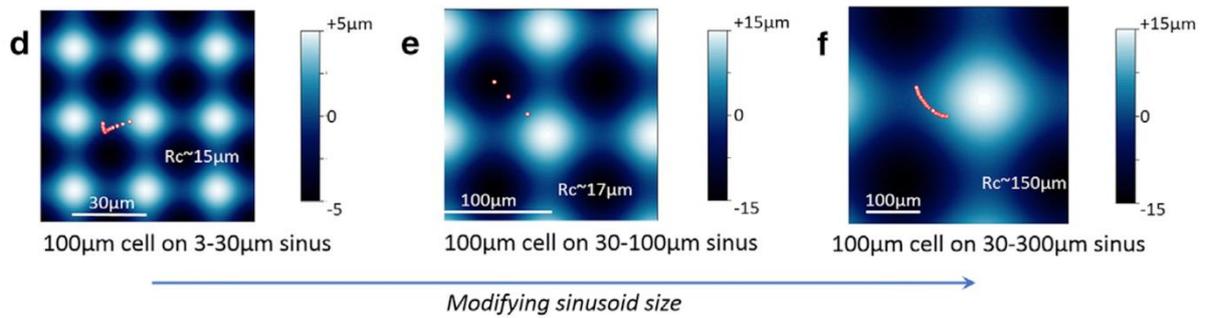

*Figure 5:* **Cell-scale curvatures optimize migration rates and persistence.** *Migration of a 100 µm cell on (a) 3-30 µm, (b) 30-100 µm, and (c) 30-300 µm sinusoids, the trajectories of the center of the cell model are indicated by the red/white data points. Cells dynamics are initiated on the flat part of the curvature (at the center of each map). Data points composing the trajectories are measured at identical time intervals, a large gap between two neighboring points indicates large migration velocity. Reproduced from [Vassaux19] (CC-BY 4.0 license).*

### 3.3. Pieces of evidence of a will of the cell to relax

We led here additional simulation of cell migration decreasing drastically cortical tension or nuclear stiffness. In both cases, the greater is the decrease, the more the model lose the capability of sensing the curvature of the substrate and migration process stopped far away from the center of a concave region. Besides as a consequence the migration velocity dropped down. So, the cell model is able to sense the curvature and to reach concave region to relax only if it possesses full integrity in its cortical tension and nucleus stiffness. Simulation results are in good agreement with in vitro observations we reported in drugged cells obtained by either blocking F-polymerization or by knocking down nucleus lamina (Fig. 6).

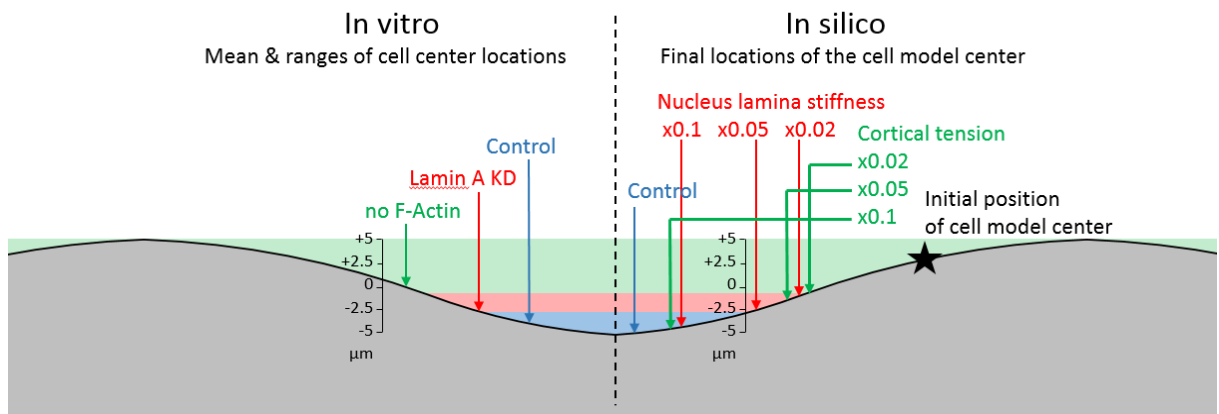

*Figure 6: In silico and in vitro results on final location of cells with altered properties of the cytoskeleton. In gray, the sinusoidal substrate. The bands in blue, red and green represent the location of the in vitro cells respectively, in control conditions, with low stiffness in the nucleus (lamin A knockdown) and without contractile cytoskeleton (no F-actin). For instance, cells in control condition located in concave area. For each in vitro condition, a vertical arrow indicates on the left side the mean position of the cells on the sinusoidal surface. On the right side, verticals arrows indicated the final position of the cell model at the end of migration in control conditions or with altered cell mechanical properties reproducing in vitro tests using drugs. In silico results are consistent with in vitro observations and lead to the same conclusion: the curvature-induced cell migration based on nucleus mechanosensitivity needs both nucleus stiffness and cytoskeleton contractility, and especially cytoskeleton contractility. Without one or both, and especially without the contractility of the cytoskeleton, cells lose their ability to detect curvature and can localize independently of the curvature gradient, whether convex or concave.*

The stiffness of the nucleus makes it an ideal topography sensor. Coupled with the actin cortex contractility, this renders a complex mechanism propelling the nucleus toward most relaxed locations inside the cell. Our *in silico* model enables to analyze and quantify directly the networks of forces established inside the cell throughout its migration (Fig. 7). This network of forces resulting from the interaction in the cytoskeleton and the nucleus is highly dynamic. Focusing on a 100µm cell migrating on the 10µm/100µm sinusoid, we observe the progressive relaxation of the forces as the cell migrates from convex to concave locations. In turn, the cell could be using its nucleus to find optimally relaxed and mechanically stable locations in its microenvironment. Concave locations in a sinusoid typically provide these two characteristics. In comparison, convex locations are highly unstable, small fluctuations in cell and nucleus centering on the topography could lead to large internal motions of organelles, highly damageable during mitosis. Flat locations are indeed more stable but do not enable the cell and its organelles to relax as much. Only cell-scale curvatures provide gradients of topography that can be perceived by the cell by impacting its mechanics. On smaller and larger radii of curvature, the topography is mostly integrated by the cell as a flat substrate, potentially not yielding a sufficiently strong mechanical signal.

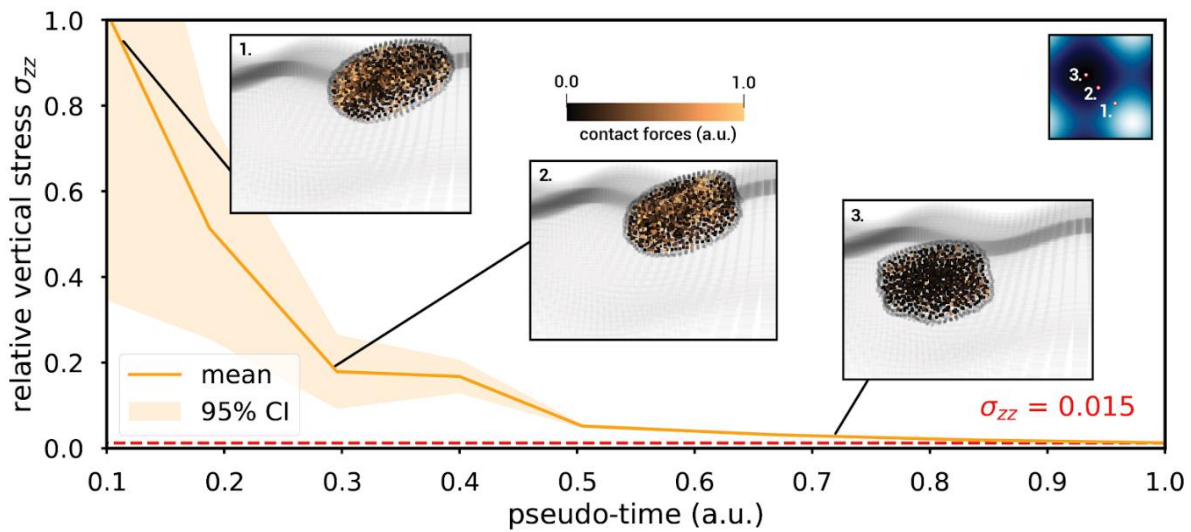

*Figure 7: **Nucleus mechanical stress relax during curvotaxis.** The mechanical vertical stress integrated over the nucleus relaxes during migration from convex to concave of a 100 µm-wide cell on a 30-100 µm sinusoid (Fig. 5.b). Snapshots of the nucleus shape and internal contact forces in the nucleoplasm are taken for three different cell positions during migration. Positions of the cell on the sinusoid in each snapshot are shown in the inset picture in the upper right corner. The vertical stress in the nucleus displays a relative decrease of 80% between onset of migration and stabilization in the nearest concave. Snapshots illustrate the simultaneous relaxation of nucleus shape, from elongated to rounded, and significant decrease of contact forces between particles constituting the nucleoplasm.*

### 3.4. Topography as a parameter of scaffold design

Properties of the topography are a significant parameter in designing scaffolds and should be chosen depending on the type of cell, more precisely the size of the cell and its nucleus, for optimal invasion and proliferation.

Our *in silico* cell model has served as a framework to integrate the hypothesized mechanism of migration called curvotaxis, whereby the cell motility is driven by the instability of its nucleus. We have been able to analyze the influence of the topography of the cell microenvironment on the motile behavior of mesenchymal stem cells. These primary results from our *in silico* stem cell model tend to show that curvotaxis could be an attempt to minimize cells mechanical energy via relaxation, as well as a way to find mechanically more stable microenvironments. Such microenvironments are beneficial for a more robust cell growth and division. Our *in silico* cell model has also enabled to quantify optimal microenvironment topographies, that is sinusoid wavelength and amplitude. Curvotaxis is rendered more efficient by sinusoids displaying cell-scale curvatures. These results could inform the design of scaffolds used in tissue engineering to promote invasion and proliferation of mesenchymal stem cells. The methodology applied in this work could be repeated for different cells types, hence enabling to design cell-type specific scaffold topographies.

## 4. Design of a new generation of biomaterials of dynamic topography aided by silico cell models

Many in vitro studies exist on the influence of topography on cell migration, however in all these works the topography remains fixed [Caballero15] (Fig. 8). We have shown that the concave regions attract the cells, but once these regions are reached, the cells stop their migration. To encourage the cells to migrate over a greater distance thanks only to the topography of the substrate and by using their curvotaxis capacity, we may propose a substrate of variable geometry, with changing topography, which can become alternately and locally concave then convex, and this cyclically. Some authors developed photochemical protocols to modulate in real time the local stiffness or strain of hydrogel substrates [Kloxin10, Chandorkar19]. We have shown that cells cultured on sinusoidal surfaces, migrate naturally, with no other stimulus than the only curvature of the surface. What would be the migratory behavior of the cells on a sinusoidal surface animated by an undulatory movement (Figure 9). Would the cells follow the ripple? Would the cells start surfing the surface of the substrate following the ripple wave? What would be the influence of the ripple frequency? The optimal frequency of ripple, or in other words the speed of the wave front, should be a priori of the same order of magnitude as the migration speed of the cell. However, what could be the influence of a ripple at very high frequencies? In such a case, a displacement of the nucleus would be observed within the cell, a displacement going in the same direction as the wave front. As we showed that the displacement of the nucleus and its decentering is a signal for the cell to migrate to center again its nucleus, this would stimulate continuous cell migration.

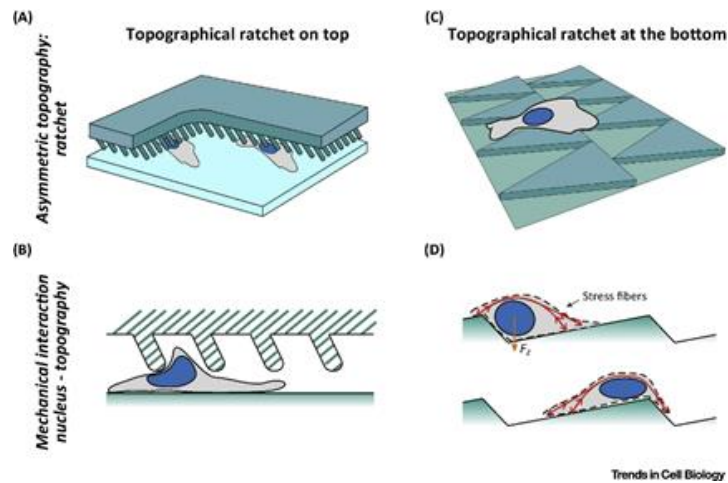

*Figure 8: Effect of the Cell Nucleus on Symmetry Breaking and Directional Migration. Cells move directionally in local asymmetric topographical ratchets imposed by confinement (A) or adhesion (C). A mechanical interaction between the cell nucleus and the tilted micropillars (B) or actomyosin stress fibers (D) guides cell polarization and motility. Reproduced from* [Caballero15] (*CC-BY 4.0 licence*).

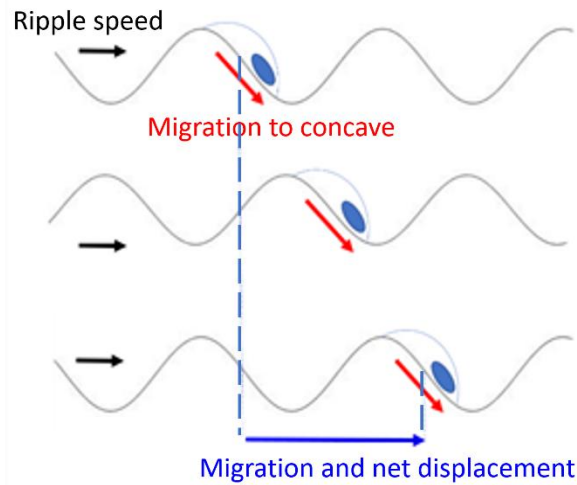

*Figure 9: Substrate animated by wave motion. Are migrating cells able to surf the wave?*

As a perspective, we can imagine an evolution of the biomaterials and scaffolds with dynamic topography to induce cell migration and invasion. Typically, in silico modelling can play a role here. Indeed, in silico experiments can be pushed beyond what is technically feasible in vitro for the time being. We propose here to analyze in silico the influence of a dynamic topography on the migratory behavior of cells using our computational cell models we presented above. In the present study, we simulated curvature-guided cell migration on a deformable sinus animated by sequential ripple motion. We imposed at the location of the cell, a deformation of the substrate to reach a 1D sinus morphology or micro-corrugated shape. Following the same process of cell migration based on the interplay between curvature-induced nucleus decentering and cell movement to center the nucleus again, we simulated iteratively the displacement of the cell until it reached the most concave region of the sinus (Fig. 10). Then the substrate deforms to become flat as at the beginning. The cell migrates with a net displacement of 45µm. Then we deformed the substrate, a second time, with the same sinusoid morphology with a dephasing of 45µm, inducing at the cell location a convex region. Following the same cell migration process, we simulate a second time cell displacement until it reached the new concave region. At the end, in imposing two deformations of the substrate, we induced cell migration in a controlled direction with a net displacement of 90µm corresponding to 1.5 times the diameter of the cell. It is worth to be noted there is no gravitation here and the cell migrates only following the nucleus decentering induced by curvature.

In the same way we can imagine to study in silico the cellular migration in interaction with a dynamic substrate, micro-channels or micro-tubes able to be piloted in radial deformation by shrinkage movements or on the contrary of swelling. In such a case, are the cells able to migrate by accompanying the deformations of the micro-tubes?

We can also design in silico, a fibrous substrate whose fibers and their crosslinking could be driven dynamically to locally animate the fibrous matrix by contraction or extension. We could experiment with the potential of migrating cells and predict whether cells are able to take advantage of the movements of their environment to migrate.

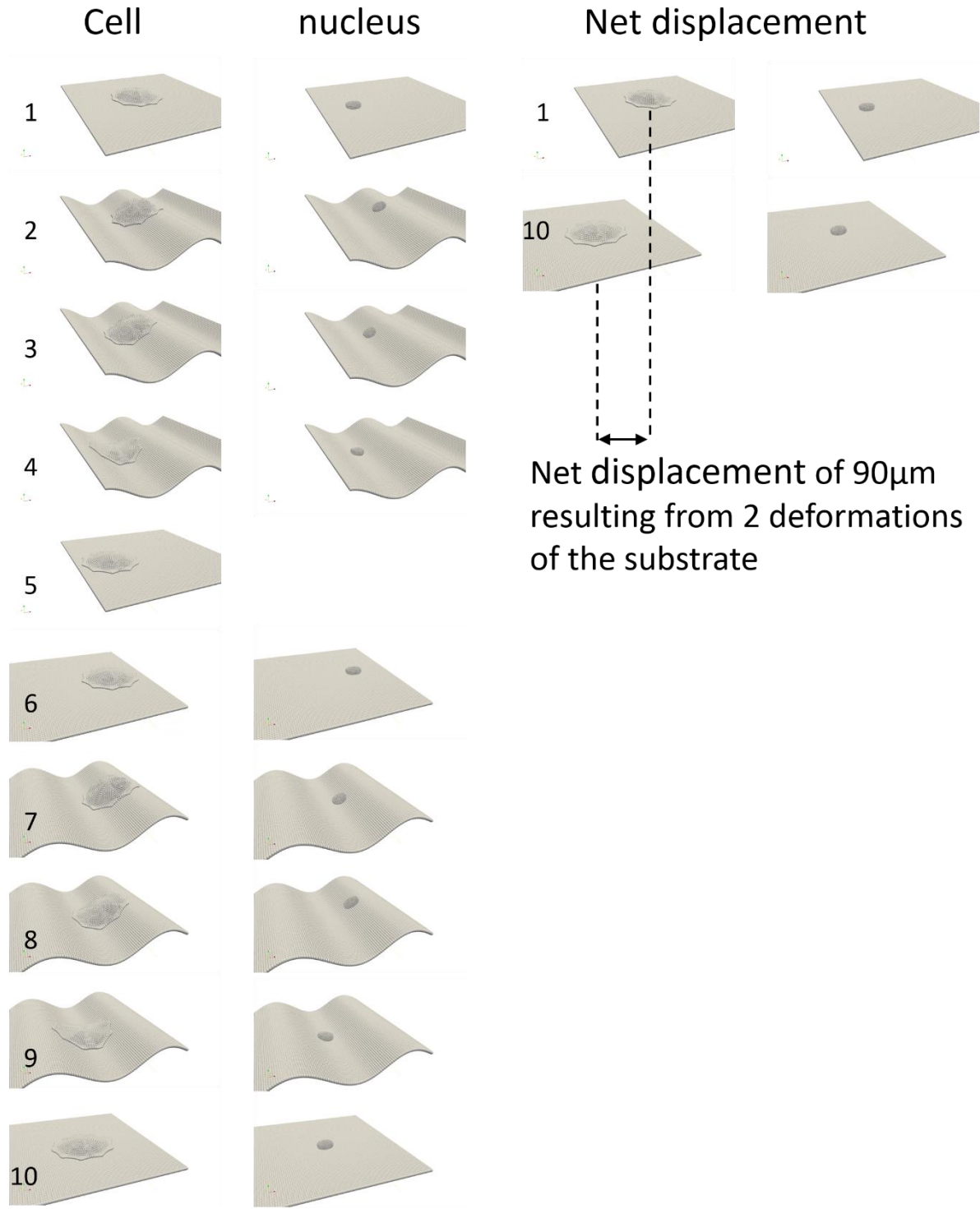

*Figure 10: image series of the sequential cell migration on a flat substrate animated by 2 successive sinus deformations. The diameter of the adherent cell shape is 100µm. After the first substrate deformation, the cell model migrated over 44.5µm. As a result, the 2nd sinus deformation was imposed with a phase shift of 44,5 µm from the first one. After the two successive deformations of the substrate, the cell model migrate over 90µm in a controlled direction. It is worth to be noted that there is no gravitation in this simulation.*

This type of dynamic topography substrates continuously stimulating cell migration could help the colonization of porous biomaterials by the cells, a colonization that is still insufficient to obtain volume tissue regeneration. And this, proposing an original method, natural because based on the normal migration of cells, and alternative to conventional mechanical methods such as perfusion or pumping of cells in suspension, methods that can damage cells. Substrates with dynamic topography could also be an alternative to biochemical methods employing chemoattractants and which raise the question of the duration of release and the duration of action.

Playing on the topography by proposing artificial and controlled geometry can make it possible to identify the processes of setting up of the adhesion and migration, to identify the cellular preferences, the processes of optimization of their form and position, of their potential adaptation, to observe the emergence of alternative solutions when one is blocked. This work could provide a great deal of information on cellular functioning and adaptation resources. This work could also inform future improvements in the design of biomaterials to stimulate migration or proliferation or differentiation by time. This cellular model could be used for the design of scaffolds specifically dedicated to bone reconstruction. To this end, the design of the scaffold should promote the invasion of mesenchymal stem cells and osteoblastic differentiation. The scaffold should also stimulate the osteoblastic activity of bone tissue synthesis via mechanical stimuli based on high apparent rigidity allowing deformation of high frequency and low amplitude. The cellular model could be a complementary approach at the cellular level to those which are developed at the tissue level and which succeed in embracing bone mechanobiology [George19, George18, Giorgio17, Lekszycki12].

In vitro experiments have their limits. While they do not fully reproduce the reality of in vivo conditions, but especially their complexity and the difficulties of producing biomaterials prevent testing many different solutions and analyze the cellular response in completely new conditions. The contribution of in silico or numerical simulation experiment, precisely allows to put in the cells situation under conditions impossible to consider in vitro and / or in vivo.

Virtually we can culture cells in a 3D environment, in contact with a material, a surface, or a fibrous matrix that would have the capability of changing its topography according to whether we are looking for the viability of stem cells by proposing a quiescent state or on the contrary a state of stress that will push them to migrate or differentiate. These controllable materials could adapt their conformation to the cellular time and specific cell function. Those smart materials are difficult or impossible to design for now. Nonetheless *in silico* experiments make it possible to overcome this problem by testing unrealistic conditions while identifying cellular behaviors never observed in vitro and dynamic microstructures and their associated deformation modes capable of stimulating cells. Based on these results, we would be able to imagine technical and feasible solutions to reproduced in vitro and in vivo the cellular response predicted by the model. The in silico approach can then join current developments in the field of intelligent materials such as meta-materials and nanomotors. For example, meta-materials thanks to their exotic electromagnetic or mechanical properties can modify their structural arrangement under the passage of electromagnetic waves or can have a negative Poisson's ratio, contracting transversely during compression [Barchiesi19; Del Vescovo14; dell'Isola19]. This type of material could be used to reproduce the optimal dynamic topography of the substrate identified by the cell model. Similarly, the properties of meta-materials could be modeled to predict and analyze the behavior of cells.